# Of Priors and Prejudice

M G Bowler

Department of Physics, University of Oxford, Keble Road, Oxford OX1 3RH, UK

**Abstract**

The methods of Maximum Entropy have been deployed for some years to address the problem of species abundance distributions. It is important to identify correctly weighting factors, or *priors*, to be applied before maximising the entropy function subject to constraints. The form of such priors depends not only on the exact problem but can also depend on the way it is set up; priors are determined by the underlying dynamics of the complex system under consideration. The problem is really one of statistical mechanics and the properties of the system yield the MaxEnt *prior* however the problem is framed. Here I calculate, in several different ways, the species abundance distribution resulting when individuals in a community are born and die independently. In the usual formulation the prior distribution for the number of species over the number of individuals is $1/n$; the problem can be reformulated in terms of the distribution of individuals over species classes. Results are obtained using master equations for the dynamics and also through the combinatoric methods of elementary statistical mechanics; the *priors* then emerge *a posteriori*. The object is not only to establish the log series species abundance distribution as the outcome of *per capita* guild dynamics but also to illuminate the nature and origin of priors. The exposition is pedagogical.

## 1. Introduction

This note is technical and unlikely to be of much interest to the general ecologist. Nonetheless it is intended to be pedagogical as well a polemic. It is concerned with the correct ways to approach the problem of relative species abundance through the techniques of statistical mechanics. That part of the physical sciences deals with very real entities (like atoms) and constraints (like conservation of energy) and while the techniques have much in common with those of maximum entropy (MaxEnt) as used in information theory, it is distinct. To my mind it is unfortunate that ecologists have encountered these techniques through probability theory, Baysian analysis and information theory, for the role of the (information theoretic) *prior* can seem mysterious. In statistical mechanics, analogues of priors have always a straightforward interpretation and are determined by physics or, here, biology. A complex system (be it a box full of gas or a forest full of trees) has a large number of accessible states which have the same gross properties; the more microstates available the more probable it is to find those gross properties. The correct way to calculate the most probable configuration depends on the properties of the complex system and that dependence translates into the appropriate *prior* if the problem is phrased in terms of MaxEnt.

Here I analyse the problem of relative species abundance in several different ways. The underlying biological principle is taken to be that species move from one abundance class to another as a result of individuals dying and giving birth independently. I formulate the problem in two ways; first by considering the most probable arrangement of species in abundance classes directly (as the problem is usually formulated using MaxEnt) and secondly by considering the distribution of the number of individuals over classes defined by all species with *n* individuals. In the first case a non-uniform *prior* is required; in the second the prior is uniform, yet the problem is the same. Both approaches are addressed through dynamical equations, the master equation approach of Volkov et al (2003, 2005) and separately through the combinatoric methods of elementary statistical mechanics. The ecological significance of the statistical mechanics of relative species abundance has been discussed at length in Pueyo et al (2007), Bowler & Kelly (2010, 2012) and is not the subject of this paper, which is intended to illuminate the nature and origins of priors. (An implicit assumption in this paper is that, for calculating species abundance distributions, species in a guild can be treated on average as having the same properties. This is treated in Pueyo et al (2007) and Bowler & Kelly (2010, 2012)).

**2. Distributing species over population classes**

We suppose that we have a set of species such that there are $s_n$ species each containing *n* individuals. The community has a fixed number of species and a fixed number of individuals, thus providing constraints to which the $s_n$ are subject as they fluctuate in response to the birth and death of individuals, immigration and even, rarely, speciation:

$$\sum_n s_n = S \tag{1a}$$

$$\sum_n n s_n = N \tag{1b}$$

In the usual approach it is supposed that before applying these constraints there exists an *a priori* probability or predilection for each species to have a population *n*, $P_\pi(n)$. The probability of finding $n_1$ individuals in species 1, $n_2$ in species 2 and so on is then proportional to the continued product $\Pi P_\pi(n_i)$, every species being counted individually. If instead of counting species individually we count the number in the class with population *n*, a particular choice of $s_n$ species with population *n* contributes a factor $P_\pi(n)^{s_n}$. There are many ways of picking a specified set of all species $\{s_n\}$ from a grand total *S* and the total number of ways is calculated simply in combinatorics (the multinomial coefficient). The weight of a specified set $\{s_n\}$ is thus given by

$$W = \frac{S! \Pi P_\pi(n)^{s_n}}{\Pi s_n!} \tag{2}$$

The most probable configuration is found by maximising $W(s_n)$ with respect to all $s_n$. It is convenient to maximise the logarithm of *W* and the constraints (1) are incorporated using the method of undetermined multipliers. Thus the function of each $s_n$ to be maximised is

$$-\ln(s_n!) + s_n \ln P_\pi(n) + \mu s_n + \lambda n s_n \qquad (3a)$$

which on applying the familiar Stirling approximation becomes

$$-s_n \ln s_n + s_n + s_n \ln P_\pi(n) + \mu s_n + \lambda n s_n \qquad (3b)$$

Differentiating with respect to $s_n$ and equating to zero the solution is

$$s_n(n) = P_\pi(n)\exp(\mu + \lambda n) \qquad (4)$$

The multipliers $\mu$ and $\lambda$ are determined from the constraints (1). If there are no constraints applied, then $\mu$ and $\lambda$ are zero and then $s_n(n) = P_\pi(n)$; the solution prior to application of constraints. (Maximising (2) is equivalent to maximising the entropy in information theory relative to a *prior* $P_\pi$.)

In real problems of the distribution of real objects over some quantity of interest, the nature of the weighting factor here denoted by $P_\pi$ is determined by the nature of the problem and can even depend on how it is framed. In the statistical mechanics of gases, it is desired to calculate the distribution of atoms over accessible energy states, subject to the constraints of conservation of number and of conservation of energy. The dynamics of atomic collisions are such that all states with a unique set of quantum numbers are equally weighted *a priori*. Atoms are distributed over these unique states according to energy and the prior probability is uniform. However, if instead of asking how many atoms are to be found in a unique quantum state of energy $E$ we were to ask how many atoms have energy $E$, the weight (or *prior*) would be the number of unique quantum states having that single energy $E$. The phenomenon is called degeneracy. In application of these notions to the problem of relative species abundance, the question is how the biological nature of the problem biases the number of species with occupation number $n$ prior to the application of constraints on the total number of species and the total number of individuals in the guild. The biological processes of birth and death can affect profoundly the chance of finding a particular species in the class labelled by $n$, because birth and death rates are (at least approximately) *per capita*. Thus a species with at one moment $n$ individuals will move into the class with $n$-1 individuals at a rate proportional to $n$; in any small interval of time all individuals have the same chance of dying and so the more there are the faster a species loses a single one. The same argument applies to individuals giving birth. Simply considering the *per capita* birth and death rates suggests that for this problem of species abundance the appropriate weighting factor (or prior) is $P_\pi(n) = 1/n$. However, though plausible verbal arguments may seem, they should be distrusted until the numbers work out. Birth and death rates appear explicitly in the master equation of Volkov et al (2003, 2005); convenient and easily understood.

### 3. The master equation for species abundance

The gain or loss of one individual in one species in the class of $n$ individuals in each species reduces the number of species in that class, $s_n$, by one. In the class there are $ns_n$ individuals each with birth and death rates $b, d$. The rate at which $s_n$ is depleted is therefore $(b + d)ns_n$. The number of species $s_n$ is augmented by a death in class $n + 1$ or by a birth in class $n - 1$. The rate of change of $s_n$ is therefore given by

$$\frac{ds_n}{dt} = -(b+d)ns_n + b(n-1)s_{n-1} + d(n+1)s_{n+1} \qquad (5)$$

For the equilibrium solution of this set of equations the right hand side of (5) is zero and

$$s_{n+1} = \frac{b}{d}\frac{n}{n+1}s_n \quad ;$$

iterating we have $s_n = \frac{1}{n}\left(\frac{b}{d}\right)^n$, apart from a normalisation constant. The second term on the right is an exponential in $n$; thus the master equation with *per capita* birth and death rates gives the solution (4) with $P_\pi(n) = 1/n$. This is the log series distribution and it is a good approximation to species abundance distributions found in nature, departures occurring at the rare species end. Thus the necessary *prior* in a MaxEnt treatment of species abundance distributions can be understood in terms of the biological processes of birth and death.

So far the discussion has differed little from that to be found in Bowler & Kelly (2010, 2012). The present paper is largely pedagogical and the material above is included for comparison with a different approach, in which individuals are distributed over species classes.

## 4. Distributing individuals over species classes

Rather than distribute numbers of species $s_n$ over classes of occupancy, suppose instead we distribute individuals over classes labelled by $n$ but such that all individuals that are members of the species in class $n$ are members of the new classes. The numbers of individuals in the new classes are $H_n = ns_n$. If the distribution of $s_n$ is log series, the distribution of $H_n$ is exponential. If the prior distribution for $s_n$ is $1/n$, then the prior for $H_n$ is uniform. The details of a complete treatment are far from obvious and are illuminating; we start by working out a master equation for the quantities $H_n$.

$H_n$ is the total number of individuals that are members of a species with population $n$. Each individual can die or give rise to a new individual. Either event removes its species from the class $n$ and thereby reduces $H_n$ by $n$ individuals. The death of a single individual in class $n+1$ adds a species to class $n$ and hence adds $n$ individuals to $H_n$. A birth in class $n-1$ similarly augments the population $H_n$ by $n$ individuals. $H_n$ changes in units of $n$ as species drop in or out of this class – this will be relevant when we consider combinatorics later. Thus we have the master equation

$$\frac{dH_n}{dt} = -n(b+d)H_n + nbH_{n-1} + ndH_{n+1} \qquad (6)$$

(The equation (6) can also be obtained by multiplying (5) by $n$ and making the general replacement $ms_m \to H_m$ but the above derivation is pedagogically useful.)

The equilibrium solution of (6) is

$$H_{n+1} = \frac{b}{d} H_n \quad ;$$

iterating we have $H_n = \left(\frac{b}{d}\right)^n$, apart from a normalisation constant. The master equation approach to relative species abundance (for *per capita* birth and death rates) suggests a uniform prior for $H_n$.

The relevant priors are obtained from the master equations by setting $b$ equal to $d$. The paper of Pueyo et al (2007) postulated that the correct prior could be obtained by requiring (spatial) scale invariance of the prior distribution. This was implemented through the minimal step of changing volume so as to remove an arbitrary single individual; invariance under this operation in fact yields the content of (5) for $b$ equal to $d$; that is, a prior for $s_n$ of $1/n$. The same invariance principle yields equations equivalent to (6) for $H_n$; that is, a uniform prior.

Regardless of whether the underlying fundamental principle is scale invariance or *per capita* rates, priors depend on how the problem is formulated but the solution to the problem does not.

**5. Combinatorics for individuals**

As in section **4** above, rather than sorting species into classes defined by the number of individuals in a species, we sort the individuals in the community into classes defined by the number of individuals in a species. We first suppose we have $N$ individuals and sort them into classes such that class 1 contains $N_1$ individuals, class 2 $N_2$ and so on. The first step is to calculate the number of ways of distributing $N$ to achieve a particular set of $N_i$. When filling class 1 with $N_1$ individuals, there are $N$ ways of choosing the first, $N$-1 ways of choosing the second ……….. to $N - N_1 + 1$ ways of choosing the last. The total number of ways is conveniently represented by the ratio of two factorial functions, $N!/(N - N_1)!$ . We then calculate the number of ways of choosing $N_2$ individuals out of the remaining $N - N_1$, which is $(N - N_1)!/(N - N_1 - N_2)!$ Continuing this process until we run out of options and multiplying the successive factors, the number of ways of achieving a particular set $N_1, N_2, \ldots$ is simply $N!$ and does not depend on the occupation numbers of the individual classes. However, if the order in which the $N_1$ individuals are chosen is of no significance, $N!$ is a gross overestimate of the number of significantly different ways of achieving our given set. The number of ways of ordering $N_1$ objects is simply $N_1!$ and if all are equivalent the number of significantly different ways of choosing $N_1, N_2, \ldots$ is the multinomial coefficient

$$M = \frac{N!}{N_1! N_2! \ldots} = \frac{N!}{\Pi N_i!} \qquad (7)$$

If the probability of finding this configuration depends on this number of different ways of achieving it, the most probable or equilibrium configuration is found by maximising $M$ with respect to all $N_i$, subject to constraints. In practice the logarithm is maximised, so we find the extremum of

$$-\sum_i \ln(N_i)! \cong -\sum_i (N_i \ln N_i - N_i) \qquad (8)$$

Maximising this function can be equivalent to maximising the Shannon entropy – MaxEnt. If $N_i$ is a function of $n_i$, constraints on $\sum_i N_i$ and on $\sum_i n_i N_i$ are incorporated by maximising

$$-N_i \ln N_i + N_i + \mu N_i + \lambda n_i N_i$$

The result is an exponential

$$N_i = \exp(\mu + \lambda n_i) \qquad (9)$$

This invites identification of the $H_n$ with the $N_i$ above, in order to achieve an exponential distribution. However, there are two things wrong and they are related. First, we have taken no account of the species structure within the $H_n$; in obtaining (7), $N_i$ could have been the number of dried peas in the $i$th urn. Secondly, one of the constraints on $N_i$, leading to (9), is wrong for $H_n$. The constraint on the total number of individuals $\sum ns_n = \sum H_n$ is not a problem, but the other constraint is on the total number of species; $\sum s_n = \sum H_n/n$. If

$$-H_n \ln H_n + H_n + \mu \frac{H_n}{n} + \lambda H_n$$

is maximised with respect to $H_n$, the result is

$$H_n = \exp(\frac{\mu}{n} + \lambda)$$

and this cannot even be normalised. The combinatorics have been formulated in a way inconsistent with the nature of the problem and the constraint on the number of species.

The mistake was that the number of equivalent ways of ordering the number of individuals $H_n$ is not $H_n!$. Suppose $n$ is equal to 2; the total number of individuals in species with 2 members is $H_2$. As soon as we choose 1 individual from among $H_2$ we have also chosen a second individual which is in the same species as our first choice. There are $H_2$ ways of choosing the first individual; the two different orderings of the two individuals in the pair are included in the $H_2$ choices. After this choice there are $H_2 - 2$ individuals left from which to select the next. This choice also entails another individual – we are selecting in pairs because of the structure of the class of all individuals in species with $n = 2$ members. The number of equivalent ways of ordering the $H_2$ individuals in this class is thus $H_2(H_2 - 2)(H_2 - 4)\dots\dots\dots\dots$

In the same way, the number of biologically equivalent ways of ordering the number $H_n$ is not $H_n!$; rather we have to make the replacement

$$H_n! \Rightarrow H_n(H_n - n)(H_n - 2n)..... = H_n!^{(n)} \qquad (10)$$

The factorial has been replaced by the multifactorial $H_n!^{(n)}$. Thus the maximum number of significant ways of achieving a given configuration, subject to the constraints on the number of individuals and the number of species, is obtained by maximising

$$-\ln H_n!^{(n)} + \mu \frac{H_n}{n} + \lambda H_n$$

The following identity holds for the multifactorial (it is easily demonstrated by extracting $n$ from each factor in (10))

$$H_n!^{(n)} = \left(\frac{H_n}{n}\right)! n^{\frac{H_n}{n}} \qquad (11)$$

[The right hand side of (11) is just the expression $s_n! n^{s_n}$ encountered in (2) with a prior (for $s_n$) of $1/n$. The substitution (10) has, in this alternative formulation, done the job of the prior in (2).]
We now have to maximise

$$-(\frac{H_n}{n} \ln \frac{H_n}{n} - \frac{H_n}{n}) - \frac{H_n}{n} \ln n + \mu \frac{H_n}{n} + \lambda H_n$$

(where we have used the Stirling approximation). Differentiating with respect to $H_n$, equating to zero and multiplying through by $n$ we have that

$$\ln H_n = \mu + \lambda n \qquad (12)$$

so that the distribution of $H_n$ is exponential in $n$ and if no constraints were applied the distribution would be uniform. The appropriate prior for $H_n = ns_n$ is uniform.

The number of ways of ordering equivalently the $H_n$ individuals is also the number of equivalent ways of extracting them all from the class defined by $n$. The extraction goes in groups in of $n$. The change in the number $H_n$ as a result of birth or death of single individuals, as expressed in the master equation (6), also occurred in groups of $n$, as species left or entered class $n$. The same underlying principles govern both the master equation and combinatoric approaches to the problem framed in terms of $H_n$.

**6. Individuals and the species abundance distribution directly**

In section **5** $N$ individuals were sorted into classes containing all individuals that are members of species with $n$ individuals. Instead, sort the $N$ individuals into bins defined by species, so that $n_i$ is the population of species $i$. The nature of the problem is expressed in the number of equivalent orderings of the $n_i$ individuals in species $i$.

This is a question inherently different from asking how many equivalent orderings there are of $n_i$ atoms in a quantum state of energy $E_i$, where the class is defined by $E_i$ and the physics is the extraction of $n_i$ as a function of $E_i$. There, the number of different equivalent ways of emptying class $i$ is certainly $n_i!$. In the species abundance problem the number of species $s_n$ in class $n$, each having $n$ members, is reduced by 1 on the removal from species $i$ of any one of its $n$ members, $n_i$ being equal to $n$ if species $i$ is a member of species class $n$. Thus for the species abundance distribution, the number of equivalent ways of removing the $n_i$ individuals from species $i$ is given by $n_i$ and not by $n_i!$; this is equivalent to the replacement (10) encountered in the previous section. Again, the number of equivalent orderings is defined by the structure of the problem. Thus the number of significantly different configurations $\{n_i\}$ is given by

$$\frac{N!}{\Pi n_i} \qquad (13)$$

There are $s_n$ species with $n$ members so switching from classifying species by taxonomy to classifying species in terms of their numbers of individuals, the combinatoric (13) becomes

$$\frac{N!}{\Pi n^{s_n}} \qquad (13a)$$

In the investigation of species abundance distributions, which species have population $n$ is immaterial. The combinatoric (13a) was obtained from a specific assignment of populations to species and so the number of ways of achieving a set $\{s_n\}$ is obtained by adding together all possible versions of (13a); that is, multiplying by the number of ways of obtaining a particular set of numbers $\{s_n\}$ from a total number of species $S$. This is given correctly by the multinomial coefficient

$$\frac{S!}{s_1! s_2! \ldots} = \frac{S!}{\Pi s_n!}$$

Thus the most probable configuration for the species abundance distribution is obtained by maximising with respect to $s_n$, subject to constraints, the function

$$W = \frac{1}{\Pi s_n! n^{s_n}}$$

and this is just equation (2) with a prior of $1/n$. Thus the combinatoric approach to the distribution of individuals over species, properly implemented, generates *a posteriori* the $1/n$ *prior* for the distribution $s_n$. The removal of any one of the $n$ individuals in a member of species class $n$ removes that species and there are $n$ equivalent ways of doing it – as in the master equation (5). This treatment embodies the same biology as the master equation, but the MaxEnt *prior* emerges from the combinatoric treatment, rather than requiring a transplant from the master equation.

## 7. Discussion

This paper has been concerned with technical aspects of the application of statistical mechanics to the problem of species abundance distributions. In applying the principles of statistical mechanics, there may be weight factors to be applied to probability functions prior to maximising the overall weight subject to constraints. There will be physical reasons for these in physics and biological reasons in biological applications. In analogous calculations using information theory and the technique of maximum entropy such factors are called *priors*. They depend on the nature of the physical or biological problem addressed and can also depend on the way in which the problem is framed. I have approached the problem of species abundance distributions using master equations incorporating *per capita* birth and death rates and used these equations to determine the appropriate priors in two different formulations of the problem. It is not necessary to employ the master equation approach; the same biology applies to combinatoric analyses. If species are put into classes of all species having *n* individuals, the prior is $1/n$; if instead individuals are sorted into classes of all individuals in species with *n* individuals, the prior is uniform. The species abundance distribution problem and its solution are the same regardless. The moral of this story is that *priors* should not be applied without thinking through the real problem and its formulation. In the problem of species abundance distributions, the right priors come out of the dynamics of the problem, or by properly deploying the same biology in combinatoric treatments. The application of an assumed invariance principle, that the prior distribution is given by invariance under changes of scale, also gets the prior right in both approaches. It is certainly true that the priors are scale invariant as implemented by Pueyo et al (2007). I would contend that this scale invariance is a result of the dynamics of *per capita* birth and death rates – which is the more fundamental is a question verging close to philosophical. The correct prior should be yielded by properly formulated principles, regardless of how the problem is posed. But if a particular ecological problem merits a particular prior $P_\pi(n)$ when framed in a particular way, there is no *a priori* reason for this particular prior to apply to a different ecological problem, however framed. *Caveat emptor!*

## References

Bowler, M.G. and C. K. Kelly, *The general theory of species abundance distributions* 2010 arXiv: 1002. 508

Bowler, M. G. and C. K. Kelly, *On the general theory of relative species abundance* 2012 Submitted to Theoretical Population Biology

Pueyo, S., F. He, and T. Zillio, *The maximum entropy formalism and the idiosyncratic theory of biodiversity.* Ecology Letters, 2007. **10**: p. 1017–1028

Volkov, I., et al., *Neutral theory and relative species abundance in ecology.* Nature, 2003. **424**: p. 1035-1037.

Volkov, I., et al., *Density dependence explains tree species abundance and diversity in tropical forests.* Nature, 2005. **438**: p. 658-661.